\documentclass[11pt]{article}

\usepackage[T1]{fontenc}
\usepackage[utf8]{inputenc}
\usepackage{lmodern}
\usepackage{textcomp}
\usepackage[margin=1in]{geometry}
\usepackage{amsmath,amssymb,bm}
\usepackage{graphicx}
\usepackage{booktabs,tabularx,array}
\usepackage[font=small,labelfont=bf,labelsep=period]{caption}
\usepackage{cite}
\usepackage{microtype}
\usepackage{xcolor}
\usepackage[colorlinks=true,linkcolor=blue!50!black,citecolor=blue!50!black,urlcolor=blue!50!black]{hyperref}

\newcommand{\um}{\,\textmu m}
\newcommand{\metric}{\int\sin^2\chi\,dz}
\newcommand{\eff}{\mathrm{eff}}
\newcommand{\dref}{d_{\mathrm{ref}}}
\newcommand{\hd}[1]{\begin{tabular}[b]{@{}c@{}}#1\end{tabular}}

\title{Unified design of short mode-evolution couplers}
\author{Shuo-Yen Tseng\\[4pt]
{\small Department of Photonics, National Cheng Kung University, Tainan, Taiwan}\\
{\small\texttt{tsengsy@mail.ncku.edu.tw}}}
\date{}

\begin{document}

\maketitle

\begin{abstract}
Short mode-evolution couplers have been designed with several seemingly different approaches, including fast quasi-adiabatic dynamics (FAQUAD), counter-diabatic (CD) driving realized by tailoring waveguide widths and spacing, CD driving realized by axis bending, and numerical optimization of the coupler geometry. Using exact two-mode coupled-mode theory in the Kramers--Henneberger frame of the bent coupler, we show that these approaches belong to the same design space. The three geometric parameters, width difference, spacing, and axis tilt, act on the light through two functions only: the coupling strength, set by the spacing, and an effective detuning, set by the width difference and the axis curvature. The gauge, bent, and mixed realizations of a CD coupler are therefore the same device and have the same bandwidth. The bandwidth is set by the angle $\chi$ between the light and the field that steers it on the Bloch sphere, integrated along the device as $\metric$, and for a CD coupler this integral is set by the reference alone. This leads to a lithography-aware design rule: make the reference as adiabatic as the minimum gap allows, by narrowing its spacing at the anticrossing, and let exact CD suppress the remaining nonadiabatic coupling. In beam propagation method (BPM) simulations on a polymer platform, a 500\um{} coupler designed this way, without numerical optimization, maintains ${>}99\%$ transfer over 274~nm at a minimum edge gap of 1.05\um, compared with 150~nm for the CD coupler of a fixed-spacing reference at nearly the same gap. Particle swarm optimization of the geometry with BPM in the loop does not exceed the design rule. Three-dimensional BPM of the BCB channel waveguides also confirms these conclusions.
\end{abstract}

\section{Introduction}

Mode-evolution couplers have been widely used in integrated optics because of their broad operational bandwidth and tolerance to fabrication imperfections~\cite{Syms1992,Xing2013,Wang2018}. They are also the building blocks of polarization splitters~\cite{Watts2005a}, rotators~\cite{Watts2005b}, and polarization splitter-rotators~\cite{Hung2024}. The underlying principle is the adiabatic following of the local eigenmodes, and the price is device length: the geometry must change slowly enough to prevent coupling between the eigenmodes. Various strategies have been proposed to shorten the devices, including minimizing unwanted mode coupling~\cite{Louisell1955,Sun2009}, optimizing shape functions~\cite{Ramadan1998,Syahriar1998}, polynomial-based designs~\cite{Ng2019}, geometry-induced suppression of crosstalk~\cite{FargasCabanillas2018}, and multi-dimensional numerical optimization~\cite{Siriani2021,Song2024}.

A second line of work draws on shortcuts to adiabaticity (STA), originally developed in quantum mechanics to speed up slow adiabatic processes~\cite{GueryOdelin2019}. Owing to the analogies between quantum systems and optical waveguides~\cite{Longhi2009}, STA methods have been adapted to coupled-waveguide devices~\cite{Chung2019,Stefanatos2014}, including invariant-based inverse engineering~\cite{Tseng2014}, fast quasi-adiabatic dynamics (FAQUAD)~\cite{MartinezGaraot2017}, and adiabaticity engineering~\cite{Wu2020}, and the path of FAQUAD through the space of coupling and detuning has been optimized~\cite{Stefanatos2026,Li2026,Liu2026}. The counter-diabatic (CD) protocol~\cite{Tseng2013,Paul2015} adds a term that cancels the nonadiabatic coupling exactly, so that light follows the eigenmodes of the original device at any length. The CD term requires an imaginary coupling coefficient, which is not available in linear waveguides, but a unitary transformation maps the protocol onto a physically realizable coupler with tailored widths and spacing~\cite{Tseng2013,Ibanez2012}. Axis bending offers a different route. Longhi showed Landau--Zener dynamics in a curved directional coupler~\cite{Longhi2005}, bent couplers were shown to be robust against wavelength and fabrication variations~\cite{Doerr2005,Morino2014,Tseng2021,Wu2022}, and we recently showed that a prescribed bending profile realizes the CD protocol in a mode-evolution coupler~\cite{Lee2026} and identified the transverse momentum as the operator that supplies the CD term~\cite{LeeTseng2026}.

These approaches were introduced separately and look different in form. Some tailor the coupling coefficient and detuning through widths and spacing, some bend the coupler axis, and some search the full geometry numerically. A similar situation existed among the STA techniques themselves, and we showed earlier that the CD protocol, invariant-based inverse engineering, and the single-shot shaped-pulse technique are equivalent in the framework of coupled-waveguide systems~\cite{Tseng2014}. Here we ask the corresponding question for the device geometry: what do width, spacing, and axis tilt do to the light, which of the published designs are the same, and what sets the bandwidth of the resulting device?

In this work, we place these approaches in one design space and show what determines their bandwidth. We first formulate an exact two-mode coupled-mode model in the Kramers--Henneberger frame, using the supermodes of the actual cross-section (Sec.~\ref{sec:model}). We then show that the three geometric parameters act on the light through two functions only, the coupling strength and an effective detuning, and describe every design by the path of the light on the Bloch sphere (Sec.~\ref{sec:knobs}). Section~\ref{sec:realizations} constructs CD couplers in the gauge, bent, and mixed realizations on a polymer platform and shows with beam propagation method (BPM) simulations that they have the same bandwidth. Section~\ref{sec:bandwidth} shows that the bandwidth follows the integral of $\sin^2\chi$ along the device, where $\chi$ is the angle between the light and the field that steers it, and that for a CD coupler this integral is set by the reference. This leads to a lithography-aware design rule (Sec.~\ref{sec:rule}): make the reference as adiabatic as the minimum gap allows, by narrowing its spacing at the anticrossing, and let exact CD suppress the remaining nonadiabatic coupling. Without numerical optimization, a 500\um{} coupler designed this way maintains ${>}99\%$ transfer over 274~nm. In Sec.~\ref{sec:pso}, particle swarm optimization (PSO) with BPM in the loop, for each geometric parameter alone and for all three, does not exceed this design. Section~\ref{sec:conclusion} discusses the place of numerical optimization and the limits of the present analysis, and Appendix~C checks the conclusions with three-dimensional BPM of the channel waveguides.

\section{Two-mode model in the Kramers--Henneberger frame}
\label{sec:model}

\subsection{Coupled-mode equation and the CD protocol}
\label{sec:cme}

\begin{figure}[tbp]
\centering
\includegraphics{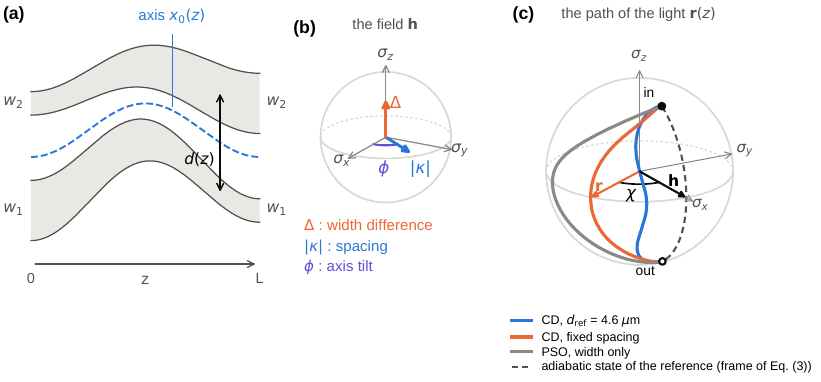}
\caption{(a) Schematic of the coupler. The waveguide widths $w_1(z)$ and $w_2(z)$, the center-to-center spacing $d(z)$, and the axis position $x_0(z)$ vary along $z$; the device length is $L$. (b) The knobs on the Bloch sphere: the width difference sets the detuning $\Delta$ ($z$ component), the spacing sets the coupling strength $|\kappa|$, and the tilt sets the phase of the coupling (rotation of the coupling vector in the $x$--$y$ plane). (c) Paths of the light $\mathbf r(z)$ on the Bloch sphere for three designs at $L=500$\um, computed from the two-mode model in the real-coupling frame of Eq.~\eqref{eq:Hz}: CD of a gap-modulated reference ($\dref=4.6$\um), CD of the fixed-spacing reference, and PSO with the width difference only (Sec.~\ref{sec:pso}). Dashed: the adiabatic state of the reference (frame of Eq.~\eqref{eq:cd}). In the real-coupling frame, the CD light keeps the latitude of this state at every $z$ and differs from it only in azimuth, by $\phi(z)$, which is zero at the ports. At $z=L/2$, the angle $\chi$ between the light $\mathbf r$ and the field $\mathbf h$ is marked for the fixed-spacing design.}
\label{fig:1}
\end{figure}

The theory described in this paper is applicable in general to weakly coupled waveguide structures described by the scalar and paraxial wave equation. We consider two waveguides, waveguide~1 and waveguide~2, placed in proximity as shown in Fig.~\ref{fig:1}(a). Their widths $w_1(z)$ and $w_2(z)$, the center-to-center spacing $d(z)$, and the position of the coupler axis $x_0(z)$ are allowed to vary along the propagation direction. Under the scalar and paraxial approximation and assuming weak coupling~\cite{Okamoto2006}, the guided-mode amplitudes in the individual waveguides are described by the coupled-mode equation
\begin{equation}
\frac{d}{dz}|\psi\rangle = -i\begin{bmatrix}\Delta & \Omega\\ \Omega & -\Delta\end{bmatrix}|\psi\rangle = -iH_0|\psi\rangle,
\label{eq:cme}
\end{equation}
where $\Omega$ is the coupling coefficient and $\Delta=(\beta_1-\beta_2)/2$ is the detuning. In the adiabatic basis defined by the mixing angle $\theta=\tan^{-1}(\Omega/\Delta)$, $U_a=\exp(-i\theta\sigma_y/2)$, the coupled-mode equation becomes
\begin{equation}
H_a = U_a^\dagger H_0 U_a - iU_a^\dagger \dot U_a = \frac{\Delta\beta}{2}\,\sigma_z + A\,\sigma_y, \qquad A = \langle\psi_+|\partial_z\psi_-\rangle = -\frac{\dot\theta}{2},
\label{eq:adiabatic}
\end{equation}
where $\Delta\beta=\beta_+-\beta_-=2(\Omega^2+\Delta^2)^{1/2}$ is the splitting of the propagation constants of the local supermodes $|\psi_\pm\rangle$, and the dot denotes $d/dz$. The second term is the nonadiabatic coupling. When it is small compared with $\Delta\beta$, light launched in $|\psi_+\rangle$ stays in $|\psi_+\rangle$, and a coupler whose $|\psi_+\rangle$ is localized in waveguide~1 at the input and in waveguide~2 at the output transfers the light between the waveguides. We quantify the adiabaticity by the parameter $C=|\dot\theta|/(\beta_+-\beta_-)$. The CD protocol adds $H_{\mathrm{cd}}=-A\sigma_y=(\dot\theta/2)\sigma_y$ to cancel the nonadiabatic coupling exactly~\cite{Tseng2013,LeeTseng2026},
\begin{equation}
H_0 + H_{\mathrm{cd}} = \Delta\,\sigma_z + \Omega\,\sigma_x + \frac{\dot\theta}{2}\,\sigma_y = \begin{bmatrix}\Delta & \Omega_s e^{-i\phi}\\ \Omega_s e^{i\phi} & -\Delta\end{bmatrix},
\label{eq:cd}
\end{equation}
with $\Omega_s=[\Omega^2+(\dot\theta/2)^2]^{1/2}$ and $\phi=\tan^{-1}(\dot\theta/2\Omega)$. Equation~\eqref{eq:cd} shows the content of the CD term in a form we use throughout: it is a coupling coefficient with a larger magnitude $\Omega_s$ and a phase $\phi$.

\subsection{Exact supermodes in the Kramers--Henneberger frame}
\label{sec:KH}

For a bent coupler, moving to the comoving coordinate $x'=x-x_0(z)$ adds the boost term $-\dot x_0\hat p_x$, with $\hat p_x=-i\partial_x$, to the paraxial propagation operator~\cite{LeeTseng2026}. The additional tilt phase of the Kramers--Henneberger (KH) transformation~\cite{Longhi2005,Lee2026} removes the boost term and replaces it by a linear transverse potential proportional to $k\ddot x_0x'$, where $k=2\pi n_s/\lambda$. We expand the field in the exact local supermodes $\psi_\pm(x';w_1-w_2,d)$ of the actual cross-section, computed by a one-dimensional finite-difference mode solver, instead of in the modes of isolated waveguides. In the KH frame, the two-mode equation in the supermode basis reads
\begin{equation}
H_a = \left(\frac{\Delta\beta}{2} - k\ddot x_0 X_z\right)\sigma_z - k\ddot x_0 X_x\,\sigma_x + A\,\sigma_y,
\label{eq:KH}
\end{equation}
where $X_z=(\langle\psi_+|x'|\psi_+\rangle-\langle\psi_-|x'|\psi_-\rangle)/2$ and $X_x=\langle\psi_+|x'|\psi_-\rangle$. Equation~\eqref{eq:KH} is the exact-supermode form of Eq.~(9) in~\cite{Lee2026}: the curvature shifts the supermode splitting through $X_z$ and couples the supermodes through $X_x$. The mixing angle is now defined by the nonadiabatic coupling of the exact supermodes, $A=-\dot\theta/2$, and reduces to $\tan^{-1}(\Omega/\Delta)$ for weak coupling. In the comoving frame, the same bend appears as the boost term, whose supermode matrix element $-\dot x_0\langle\psi_+|\partial_x\psi_-\rangle$ is purely $\sigma_y$~\cite{LeeTseng2026}. The two forms are related by integration by parts and by the identity $\langle\psi_+|\partial_x\psi_-\rangle=k\,\Delta\beta\,X_x$, and they agree to first order in the tilt~\cite{LeeTseng2026}.

\section{Three knobs, two functions, and the path of the light}
\label{sec:knobs}

\subsection{A tilt turns the phase of the coupling}
\label{sec:tilt}

\begin{figure}[tbp]
\centering
\includegraphics[width=0.3\textwidth]{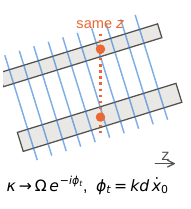}
\caption{Phase fronts across two tilted waveguides separated by $d$; the coupling acquires the phase $\phi_t=kd\dot x_0$.}
\label{fig:2}
\end{figure}

When the coupler axis is tilted by $\dot x_0$, the phase front of the light is tilted with it. The two waveguides, separated by $d$, then see phase fronts that differ by $\phi_t=kd\dot x_0$ [Fig.~\ref{fig:2}], and the coupling coefficient between them acquires this phase:
\begin{equation}
H = \Delta\,\sigma_z + \Omega\left(\cos\phi_t\,\sigma_x + \sin\phi_t\,\sigma_y\right) = \begin{bmatrix}\Delta & \Omega e^{-i\phi_t}\\ \Omega e^{i\phi_t} & -\Delta\end{bmatrix}.
\label{eq:tilt}
\end{equation}
Comparing Eq.~\eqref{eq:tilt} with Eq.~\eqref{eq:cd}, the tilt supplies the phase of the coupling coefficient that the CD protocol asks for, but not its magnitude. The tilt does not change the magnitude of the coupling, which is set by the spacing alone. The CD term, however, also increases the magnitude, from $\Omega$ to $\Omega_s=[\Omega^2+(\dot\theta/2)^2]^{1/2}$ in Eq.~\eqref{eq:cd}, because it adds a coupling component at right angles to $\Omega$. A CD coupler therefore needs a narrower spacing than its reference wherever $\dot\theta\neq0$, whatever share of the phase the tilt supplies: the bent realization narrows the gap as much as the gauge realization (Secs.~\ref{sec:three} and~\ref{sec:drc}). The width difference sets $\Delta$. The three geometric knobs therefore act on three quantities, the magnitude of the coupling (spacing), its phase (tilt), and the detuning (width difference).

\subsection{A frame rotation moves the coupling phase into the detuning}
\label{sec:frame}

The coupling phase is not independent of the detuning. Applying the $z$-axis rotation $U_z=\exp(-i\phi\sigma_z/2)$ used in the gauge realization of CD driving~\cite{Tseng2013} to a coupling coefficient $\kappa=|\kappa|e^{-i\phi}$, of magnitude $|\kappa|$ and phase $\phi$, gives
\begin{equation}
H_z = U_z^\dagger H U_z - iU_z^\dagger \dot U_z = |\kappa|\,\sigma_x + \Delta_{\eff}\,\sigma_z, \qquad \Delta_{\eff} = \Delta - \frac{\dot\phi}{2}.
\label{eq:Hz}
\end{equation}
$U_z$ only changes the phases of the waveguide amplitudes, so the power in each waveguide is the same in both frames. The light therefore responds to two functions only: the coupling strength $|\kappa|(z)$, set by the spacing, and the effective detuning $\Delta_{\eff}(z)$. For a tilt at fixed spacing, $\dot\phi/2=kd\ddot x_0/2$, so the axis curvature acts as a detuning, as in Longhi's curved coupler~\cite{Longhi2005} and in the curvature terms of Eq.~\eqref{eq:KH}. The same $\Delta_{\eff}$ can be produced by the width difference alone, by the curvature alone, or by any mixture. For the CD protocol of Eq.~\eqref{eq:cd}, Eq.~\eqref{eq:Hz} gives $|\kappa|=\Omega_s$ and $\Delta_{\eff}=\Delta-\dot\phi/2$, the straight-coupler realization of~\cite{Tseng2013}; supplying $\phi$ by the tilt instead gives the bent realization of~\cite{Lee2026,LeeTseng2026}. Both are the same effective Hamiltonian. In both, $|\kappa|=\Omega_s$ is supplied by the spacing; the realization decides only whether the detuning $-\dot\phi/2$ comes from the width difference or from the axis curvature.

\subsection{Path on the Bloch sphere}
\label{sec:bloch}

The two-mode evolution has a simple geometric form. Writing $H=\mathbf h\cdot\bm\sigma$ and the state as a Bloch vector $\mathbf r(z)$, Eq.~\eqref{eq:cme} gives $\dot{\mathbf r}=2\mathbf h\times\mathbf r$. The polar angle of $\mathbf r$ sets the power split between the waveguides and its azimuth their relative phase, so $\mathbf r(z)$ is the path of the light through the device [Fig.~\ref{fig:1}(c)]. Taking $\mathbf r\times$ of both sides and using $|\mathbf r|=1$ gives $\mathbf r\times\dot{\mathbf r}=2[\mathbf h-(\mathbf h\cdot\mathbf r)\mathbf r]$. Any field $\mathbf h$ that drives a given path $\mathbf r(z)$ can therefore be split as
\begin{equation}
\mathbf{h} = (\mathbf{h}\cdot\mathbf{r})\,\mathbf{r} + \frac{1}{2}\,\mathbf{r}\times\dot{\mathbf{r}}.
\label{eq:split}
\end{equation}
The first term is a field whose eigenstate is the actual state of the light; it does not move the light and adds only a phase; the second moves the light along the path: it is the CD term $\frac12(\mathbf r\times\dot{\mathbf r})\cdot\bm\sigma$ of a reference whose adiabatic state is the light itself~\cite{Berry2009}, which for the reference of Eq.~\eqref{eq:cd}, $\mathbf r=(\sin\theta,0,\cos\theta)$, is $(\dot\theta/2)\sigma_y$. This is the waveguide counterpart of the multiple Schr\"odinger pictures of STA~\cite{Ibanez2012}. A device is therefore fixed by its path. In the real-coupling frame of Eq.~\eqref{eq:Hz}, $\mathbf h$ has no $y$ component, which fixes the first term once the path is chosen, so designing a coupler means choosing a path from the input state to the output state that the available coupling strength and detuning can drive.

The relative size of the two terms is the angle $\chi$ between the light and the field [Fig.~\ref{fig:1}(c)]. A rotation about $z$ changes only the relative phase of the waveguide amplitudes and adds a detuning (Sec.~\ref{sec:frame}), so we measure $\chi$ in the frame that follows the relative phase of the light. In that frame the path stays in a plane through the $z$ axis, the second term of Eq.~\eqref{eq:split} is normal to that plane, and
\begin{equation}
\tan\chi = \frac{|\dot\Theta|}{2h_\parallel},
\label{eq:tanchi}
\end{equation}
where $\Theta$ is the polar angle of $\mathbf r$, which sets the power split between the waveguides, and $h_\parallel=\mathbf h\cdot\mathbf r$ is the first term of Eq.~\eqref{eq:split} in that frame. $\chi$ does not change under rotations about $z$ and is therefore the same for the gauge, bent, and mixed realizations. For $\chi=0$, the light follows the local eigenstate of the field; for $\chi=90^\circ$, the field is perpendicular to the light, which precesses about it as in a directional coupler.

For a CD design, $\chi$ is known before the design is made. Its path is the adiabatic state of the reference, and the frame that follows its relative phase is the frame of Eq.~\eqref{eq:cd}. There $h_\parallel=(\beta_+-\beta_-)/2$ and $\Theta=\theta$, so that
\begin{equation}
\tan\chi = \frac{|\dot\theta|}{\beta_+ - \beta_-} = C,
\label{eq:tanchiC}
\end{equation}
where $C$ is the adiabaticity parameter of the reference. A reference that is nearly adiabatic gives a small angle, and the light stays close to the local eigenstate. In the real-coupling frame of Eq.~\eqref{eq:Hz}, the light of a CD coupler keeps the latitude of the adiabatic state of the reference at every $z$ and differs from it only in azimuth, by the coupling phase $\phi(z)$ of Eq.~\eqref{eq:cd}, which is zero at the ports [Fig.~\ref{fig:1}(c)].

Designs therefore differ in how they choose the path and in how they realize it. A CD coupler takes the adiabatic state of a chosen reference as its path, and a frame change that reduces to the identity at the ports~\cite{Ibanez2012} deforms the path and keeps its end points. A rotation about $z$, Eq.~\eqref{eq:Hz}, only decides whether the coupling phase is supplied by the width difference (gauge), the tilt (bend), or a mixture, and changes neither the path nor $\chi$. We therefore expect the bandwidth to be set by the path, chosen through the reference, and to be insensitive to the realization. Sections~\ref{sec:realizations} and~\ref{sec:bandwidth} test this.

Table~\ref{tab:1} lists published approaches by how they choose the path and how they realize it. Each is a rule for choosing the path; none lies outside this description.

\begin{table}[tbp]
\caption{Published approaches as choices of the path of the light and of its realization.}
\label{tab:1}
\centering
\small
\renewcommand{\arraystretch}{1.15}
\begin{tabularx}{\textwidth}{>{\raggedright\arraybackslash}p{0.25\textwidth}>{\raggedright\arraybackslash}X>{\raggedright\arraybackslash}p{0.27\textwidth}}
\toprule
Approach & How the path is chosen & Realization\\
\midrule
Adiabatic tapers~\cite{Syms1992,Xing2013,Wang2018,Louisell1955,Sun2009,Ramadan1998,Syahriar1998,Ng2019}, multi-parameter adiabatic design~\cite{Siriani2021} & the local eigenstate of the device, swept slowly enough & width difference and spacing; width and axis position at fixed gap~\cite{Siriani2021}\\
FAQUAD~\cite{MartinezGaraot2017}, adiabaticity engineering~\cite{Wu2020} & the local eigenstate, swept at a uniform or engineered adiabaticity & width difference and spacing\\
Curved couplers~\cite{Longhi2005} & the local eigenstate, swept by a curvature-driven detuning & curvature\\
CD by width and spacing~\cite{Tseng2013} & the adiabatic state of a reference, followed exactly & width difference and spacing (rotation about $z$)\\
CD by bending~\cite{Lee2026,LeeTseng2026} & the same path & tilt for the coupling phase, spacing for its magnitude\\
Invariant-based engineering, robust STA~\cite{Tseng2014} & the eigenstate of a dynamical invariant, chosen for error tolerance & width difference and spacing\\
Multiple Schr\"odinger pictures~\cite{Ibanez2012} & frame changes that keep the end points of a path & any\\
Geometry-induced crosstalk suppression~\cite{FargasCabanillas2018}, coefficient canceling~\cite{Siriani2021}, fast quasi-adiabatic couplers~\cite{Song2024} & the tilt cancels the coupling between the local modes at fixed gap: the CD term to first order in the tilt, without the larger coupling magnitude it also needs (Sec.~\ref{sec:tilt}); the light leaves the local eigenstate and returns; in~\cite{Song2024}, the tilt is refined by PSO with FDTD in the loop & width taper and axis tilt at fixed gap\\
Path-optimized FAQUAD~\cite{Stefanatos2026,Li2026,Liu2026} & the local eigenstate along the path of smallest adiabaticity integral, on the boundary of the allowed coupling and detuning, swept at uniform adiabaticity & coupling and detuning~\cite{Stefanatos2026,Liu2026}; width difference and axis shift at fixed gap~\cite{Li2026}, as in~\cite{FargasCabanillas2018,Siriani2021}\\
Numerical optimization~\cite{Song2024} and Sec.~\ref{sec:pso} & implicit: the path the search converges to & axis tilt at fixed gap~\cite{Song2024}; all knobs (Sec.~\ref{sec:pso})\\
\bottomrule
\end{tabularx}
\end{table}

\section{One coupler, three realizations}
\label{sec:realizations}

\subsection{Platform and simulation}
\label{sec:platform}

We use the polymer waveguide platform of~\cite{Lee2026}, the effective-index model of the BCB channel waveguides of~\cite{Tseng2013}. The cladding index is $n_s=1.5$ and the core index difference is 0.016 at $\lambda=1.55$\um. The sum of the waveguide widths is fixed at $w_1+w_2=6.8$\um, and the width difference $\delta w=w_1-w_2$ changes from $+1.6$\um{} at the input ($W_1=4.2$\um, $W_2=2.6$\um) to $-1.6$\um{} at the output. The center-to-center spacing at the ports is $d_0=5.5$\um, and we call $d-(w_1+w_2)/2=d-3.4$\um{} the edge gap. The untreated reference is a smoothstep taper $\delta w=1.6[1-2u(z/L)]$\um, $u(s)=3s^2-2s^3$, at fixed spacing; we call it the smoothstep reference. We perform the BPM simulations of Secs.~\ref{sec:realizations}--\ref{sec:pso} using a scalar 2D Crank--Nicolson BPM solver based on the finite-difference approach ($\Delta x=0.03$\um, $\Delta z=0.5$\um) with a 4\um{} absorbing layer at the edges of a 48\um{} window~\cite{Kawano2004}. Material dispersion is not included. The figure of merit is the supermode fidelity $P_+=|\langle\psi_+(L)|E(L)\rangle|^2$ for light launched in $\psi_+$ at the input, as in~\cite{LeeTseng2026}. To compare tolerances, we also record the worst $P_+$ over an error range, with the errors applied one at a time: wavelength 1.50--1.60\um, common width error $\pm50$~nm, spacing error $\pm100$~nm, and antisymmetric width error $\pm25$~nm. At $L=500$\um, the smoothstep reference alone reaches $P_+=0.51$. We characterize the bandwidth by BW99, the wavelength range over which $P_+>0.99$, within a sweep from 1.40 to 1.70\um. The two-mode model of Eq.~\eqref{eq:KH} reproduces BPM closely (Appendix~A); all transfer and bandwidth values below are from BPM.

\subsection{Three CD couplers of the smoothstep reference}
\label{sec:three}

We now build CD couplers of the smoothstep reference at $L=500$\um. The goal is three devices that realize the same CD Hamiltonian in different ways: with the width difference and the spacing alone (the gauge realization, $f=0$)~\cite{Tseng2013}, with the coupling phase supplied by the axis tilt (the bent realization, $f=1$)~\cite{Lee2026}, and with an even mixture ($f=0.5$). Each device is specified by its geometry $\delta w(z)$, $d(z)$, and $x_0(z)$, which we construct directly from the exact coefficients of Eq.~\eqref{eq:KH}, without fitting relations between geometry and coupled-mode parameters (Appendix~B). The mixing angle $\theta$ of the reference is integrated along its path through the $(\delta w,d)$ plane, the CD Hamiltonian of Eq.~\eqref{eq:cd} is brought to a real coupling by the rotation about $z$ of Eq.~\eqref{eq:Hz}, the tilt supplies a fraction $f$ of the coupling phase, and the width difference and the spacing supply the rest. The construction takes about 0.1~s per design.

\begin{figure}[tbp]
\centering
\includegraphics{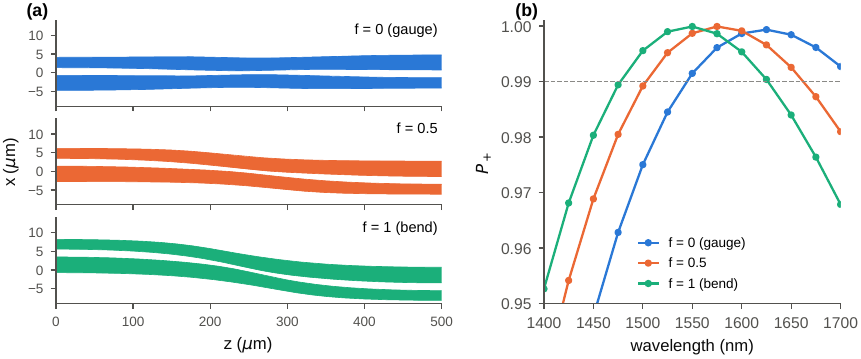}
\caption{CD couplers of the smoothstep reference at $L=500$\um{} in the gauge ($f=0$), mixed ($f=0.5$), and bent ($f=1$) realizations. (a) Waveguide geometries. (b) $P_+$ against wavelength.}
\label{fig:3}
\end{figure}

Figure~\ref{fig:3} shows the result. The gauge ($f=0$), the mixture ($f=0.5$), and the bent design ($f=1$) look different [Fig.~\ref{fig:3}(a)]: the gauge design uses the width difference alone, while the bent design uses the axis curvature for most of the detuning. All three narrow the spacing at the anticrossing equally, from 5.5 to 4.46\um, because the coupling strength $|\kappa|=\Omega_s$ is set by the spacing alone (Sec.~\ref{sec:tilt}). The bent design reaches $P_+=0.9999$ in BPM, the mixture 0.9987, and the gauge design 0.9915. The difference is at the ports. The smoothstep has $\ddot\theta\neq0$ there, so the CD term changes the effective detuning at the ports; the bend supplies this change by curvature and keeps the port geometry of the reference, while the gauge design must change the port width difference to 1.26\um, and its port supermodes differ from those of the reference by 0.11~rad in mixing angle. The mismatches at the two ports interfere, which also shifts its band toward longer wavelengths. Launched in and projected on the port supermodes of the reference, the gauge design reaches $P_+=0.9994$. Their bandwidths, BW99 $=155$, 159, and 149~nm for $f=0$, 0.5, and 1, agree [Fig.~\ref{fig:3}(b)]. The bent design keeps the port geometry but starts with a finite curvature at the ports. This curvature is part of the CD Hamiltonian; because $\dot x_0=0$ at the ports, the KH frame coincides there with the lab frame, and the supermodes of the straight port cross-sections are the input and output states. A reference with $\ddot\theta=0$ at the ports, such as the quintic smoothstep, keeps the port geometry for every realization, but it is steeper mid-device and needs a 0.25\um{} narrower gap.

The realization is an engineering choice, for example to keep the port geometry fixed or to limit the curvature, with little effect on the bandwidth.

\section{What sets the bandwidth}
\label{sec:bandwidth}

\begin{figure}[tbp]
\centering
\includegraphics{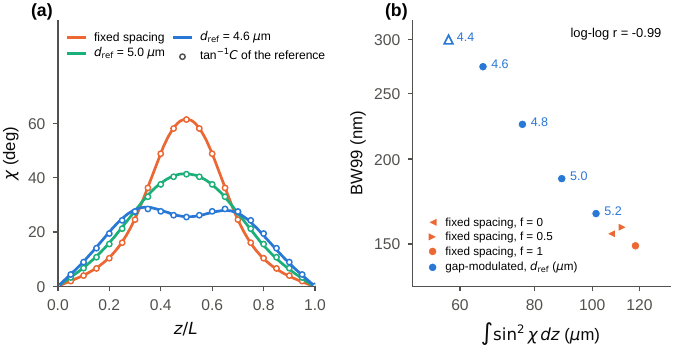}
\caption{(a) Angle $\chi$ between the light and the field along $z$ for the exact CD couplers of the fixed-spacing reference and of gap-modulated references with $\dref=5.0$ and 4.6\um{} at $L=500$\um{} (lines), computed from the two-mode model; circles: $\tan^{-1}C$ of the reference, with $C=|\dot\theta|/(\beta_+-\beta_-)$ its adiabaticity parameter [Eq.~\eqref{eq:tanchiC}]. (b) BW99 against $\metric$ for the CD couplers of the fixed-spacing reference in the gauge, mixed, and bent realizations ($f=0$, 0.5, and 1) and of the gap-modulated references of Table~\ref{tab:2} ($f=1$, labeled by $\dref$ in \textmu m); the triangle reaches the 300~nm edge of the sweep.}
\label{fig:4}
\end{figure}

Section~\ref{sec:realizations} showed that the realization does not set the bandwidth. To see what does, we use the angle $\chi$ of Sec.~\ref{sec:bloch}. Figure~\ref{fig:4}(a) shows $\chi$ along the CD couplers of the smoothstep reference and of two references whose spacing is narrowed mid-device (Sec.~\ref{sec:rule}). $\chi$ is largest at the anticrossing, where the reference is least adiabatic, and the narrower spacing lowers it. Computed from the light in the two-mode model, $\chi$ agrees with $\tan^{-1}C$ of the reference to within $0.8^\circ$ along the whole device [Fig.~\ref{fig:4}(a), circles].

Why $\chi$ sets the bandwidth can be read from the path. A field at angle $\chi$ from the light rotates it about the field direction on a circle of radius $\sin\chi$. The design makes this rotation exactly right at the design wavelength, so that the light arrives at the output state. Away from the design wavelength, the coupling and the detuning change, the accumulated rotation $\int(\beta_+-\beta_-)\,dz$ changes by some $\delta\Phi$, and the light misses the output state by about $\delta\Phi\sin\chi$ on the sphere. For $\chi\to0$, the light sits on the rotation axis and follows the field whatever the rotation angle; this is why adiabatic transfer is broadband. For $\chi$ near $90^\circ$, the light moves on a great circle, as in a directional coupler, and the error is largest.

The miss is made wherever the light is away from the field, and it adds up along the device. A change $\delta$ of the precession rate turns the light about the field by $\delta\,dz$ on a circle of radius $\sin\chi$. The rest of the device carries this displacement to the output and turns it by the remaining rotation angle, so the contributions add as phasors. To first order in $\delta$,
\begin{equation}
1 - F \approx \frac{1}{4}\left|\int_0^L \delta\,\sin\chi\,e^{i\xi}\,dz\right|^2,
\label{eq:firstorder}
\end{equation}
where $\xi(z)$ is the direction, in the plane tangent to the sphere at the output state, into which the rest of the device turns the displacement made at $z$. Over a band of wavelengths, where the rotation angle changes, the contributions lose their relative phase and add in power. The loss then grows with $\int\delta^2\sin^2\chi\,dz$, which for a nearly uniform $\delta$ is proportional to $\metric$, the integral of $\sin^2\chi$ along the device; $\sin^2\chi$ is the variance of the normalized field $\hat{\mathbf h}\cdot\bm\sigma$ in the state of the light, that is, how far the light is from an eigenstate of the field. This is an estimate, not a bound, and we test it below. For a uniform $\chi$, the sum can be done exactly: at constant adiabaticity, $1-F=\sin^2\chi\,\sin^2(\Phi/2)$, with $\Phi$ the total rotation angle~\cite{Stefanatos2026}.

For a CD coupler, $\tan\chi=C$ [Eq.~\eqref{eq:tanchiC}], so the metric, $\metric=\int C^2/(1+C^2)\,dz$, can be read from the reference alone, before the coupler is built. Figure~\ref{fig:4}(b) shows that it orders the bandwidth of the CD couplers. The gauge, mixed, and bent realizations of the smoothstep reference have nearly the same integral, 108--118\um, and the same bandwidth. Narrowing the spacing of the reference at the anticrossing lowers the integral to 57\um{} and raises BW99 from 150~nm to the 300~nm edge of the sweep. For the seven designs below the edge, the correlation coefficient between $\log$ BW99 and $\log\metric$ is $-0.99$. A CD coupler is therefore as broadband as its reference is adiabatic, whatever the realization.

\section{Design rule}
\label{sec:rule}

Section~\ref{sec:bandwidth} showed that the bandwidth falls as $\metric$ grows and that, for a CD coupler, this integral is set by its reference. The smoothstep reference at fixed spacing is far from adiabatic at $L=500$\um{} ($C$ up to 1.84), and its CD coupler reaches BW99 $=150$~nm. The coupling strength sets how adiabatic a reference can be at a given length, and the spacing is the only knob on it (Sec.~\ref{sec:knobs}). These results point to a design rule: make the reference as adiabatic as the minimum gap allows, by narrowing its spacing at the anticrossing, and let exact CD suppress the remaining nonadiabatic coupling. We therefore change the spacing of the reference itself, not only that of the CD correction.

\subsection{Gap-modulated reference}
\label{sec:gapref}

Within a fixed length, the reference becomes more adiabatic when the coupling is increased at the anticrossing. For small $C$, $\metric\approx\int C^2\,dz\geq J^2/L$, where $J=\int C\,dz=\int|d\theta|/(\beta_+-\beta_-)$ is the adiabaticity integral of the reference~\cite{Stefanatos2026}. $J$ depends only on the path of the reference through the $(\delta w,d)$ plane, and the path of smallest $J$ runs along the boundary of the allowed parameters, here the minimum spacing~\cite{Stefanatos2026}. We therefore keep the smoothstep width taper and narrow the spacing of the reference mid-device,
\begin{equation}
d(z) = d_0 - (d_0 - \dref)\,\sin^2(\pi z/L),
\label{eq:gap}
\end{equation}
which moves the path toward this boundary mid-device, where the width difference changes fastest, and leaves the ports unchanged. We call it the $\sin^2$ profile. The mixing angle is integrated along the $(\delta w,d)$ path, and the exact CD design follows as in Appendix~B. The realization now matters at the ports. The CD term changes the detuning near the ports; the bend supplies this change by curvature and keeps the port geometry of the reference, while the gauge realization must change the port widths and fails for boosted references ($P_+=0.84$--0.99). We therefore use the bend throughout, $f=1$. The CD correction narrows the gap further than the reference, so the relevant fabrication number is the minimum edge gap of the final device, not $\dref$.

\begin{table}[tbp]
\caption{Exact CD couplers of the smoothstep reference with the $\sin^2$ gap profile of Eq.~\eqref{eq:gap} at $L=500$\um, bent realization ($f=1$); $\dref=5.5$\um{} is the fixed-spacing reference. Edge gap is the minimum edge-to-edge gap of the final device. Worst over errors is the lowest $P_+$ over the error range of Sec.~\ref{sec:platform}. $\metric$ is the bandwidth metric of Sec.~\ref{sec:bandwidth}. BW99 marked $\geq300$~nm reaches the edge of the 1.40--1.70\um{} sweep.}
\label{tab:2}
\centering
\small
\begin{tabular}{lccccc}
\toprule
$\dref$ (\textmu m) & \hd{Edge gap\\(\textmu m)} & $P_+$ & \hd{BW99\\(nm)} & \hd{Worst over\\errors} & \hd{$\metric$\\(\textmu m)}\\
\midrule
5.5 (fixed spacing) & 1.06 & 0.99991 & 150 & 0.9917 & 117\\
5.2 & 1.18 & 0.99978 & 166 & 0.9926 & 101\\
5.0 & 1.19 & 0.99967 & 187 & 0.9931 & 89\\
4.8 & 1.15 & 0.99962 & 225 & 0.9938 & 76\\
\textbf{4.6 (design-rule coupler)} & 1.05 & 0.99957 & 274 & 0.9948 & 66\\
4.4 & 0.92 & 0.99949 & $\geq300$ & 0.9959 & 57\\
\bottomrule
\end{tabular}
\end{table}

\subsection{Design-rule coupler}
\label{sec:drc}

\begin{figure}[tbp]
\centering
\includegraphics{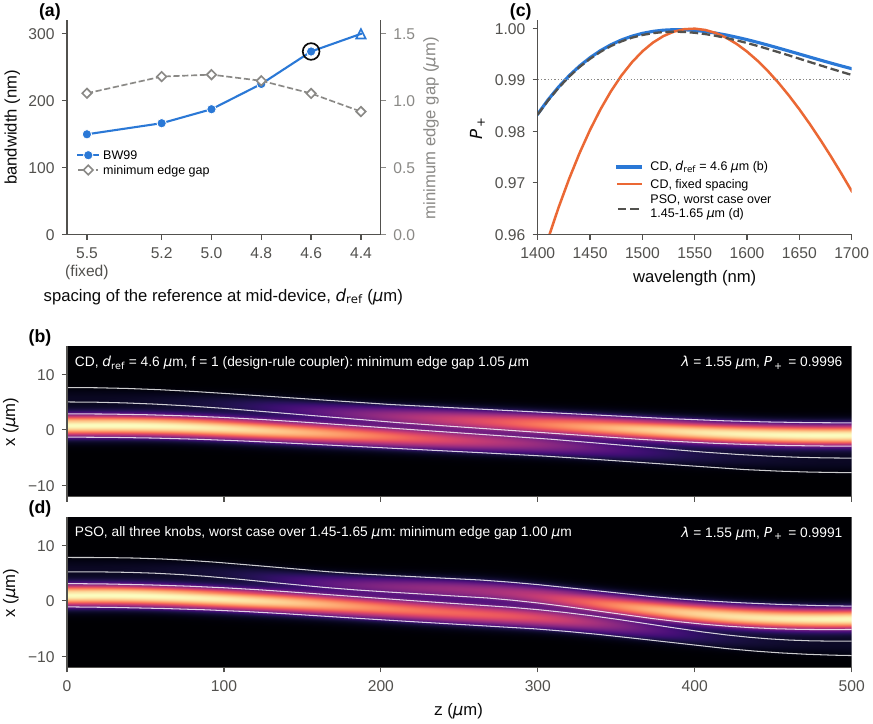}
\caption{(a) BW99 (left axis) and minimum edge gap of the device (right axis) of the CD couplers of Table~\ref{tab:2} against the mid-device spacing of the reference $\dref$; $\dref=5.5$\um{} is the fixed-spacing reference. Circle: the design-rule coupler of (b) and (c); triangle: reaches the edge of the sweep. (b) BPM light evolution at 1.55\um{} in the design-rule coupler ($L=500$\um, $\dref=4.6$\um, $f=1$), with the waveguide cores outlined. (c) $P_+$ against wavelength of the design-rule coupler, the CD design of the fixed-spacing reference, and the PSO design of (d). (d) BPM light evolution at 1.55\um{} in the best robust PSO design (all three knobs, worst $P_+$ over 1.45--1.65\um; Sec.~\ref{sec:pso} and Table~\ref{tab:3}), with a minimum edge gap of 1.00\um.}
\label{fig:5}
\end{figure}

Figure~\ref{fig:5}(a) shows the bandwidth and the minimum edge gap of these designs. Narrowing the reference spacing from 5.5 to 4.6\um{} raises BW99 from 150 to 274~nm, while the minimum edge gap of the device stays between 1.05 and 1.19\um. The CD correction narrows the gap below that of the reference (Sec.~\ref{sec:tilt}), by 1.05\um{} for the fixed-spacing reference but only by 0.15\um{} for $\dref=4.6$\um, because a more adiabatic reference needs a smaller correction. Only below $\dref=4.6$\um{} does the minimum gap fall below that of the fixed-spacing design. We take $\dref=4.6$\um{} for the design-rule coupler [Figs.~\ref{fig:5}(b) and~\ref{fig:5}(c)] because its minimum edge gap, 1.05\um, matches that of the CD coupler of the fixed-spacing reference (1.06\um), so that the two are compared at the same lithography. It maintains $P_+>0.99$ over 274~nm, and its worst $P_+$ over the error range is 0.9948. The design needs no optimizer, and every part of its geometry has a physical reading: the gap modulation makes the reference adiabatic, and the remaining CD correction is small. The gain comes from the reference, not from the realization.

A longer device relaxes the lithography: with the same profile, BW99 $>260$~nm is reached at a minimum edge gap of 1.26\um{} for $L=700$\um{} and 1.48\um{} for $L=1000$\um.

\section{Numerical search}
\label{sec:pso}

To test the design rule against designs found without a model, we optimize the coupler with BPM in the loop using particle swarm optimization (PSO)~\cite{Kennedy1995}. Each varied profile is that of the smoothstep reference plus six sine modes that vanish at the ports: $\delta w(z)=\delta w_{\mathrm{ref}}(z)+\sum_m b_m\sin(m\pi z/L)$, and likewise for $d(z)-d_0$ and $\dot x_0(z)$, with $m=1,\ldots,6$. The port geometry is therefore that of the reference, and each varied knob adds six parameters. The bounds of the coefficients decrease as $1/m$, from 1\um, 2.5\um, and 0.06~rad at $m=1$, to favor smooth profiles, and the geometry is constrained to $|\delta w|\leq2.1$\um, $4.4\text{\um}\leq d\leq8.3$\um{} (minimum edge gap 1.0\um), and $|\dot x_0|\leq0.08$~rad. A swarm of 36 particles runs for 200 iterations, 7200 BPM simulations per optimization, and every case is repeated with three random seeds (five for all three knobs).

\begin{figure}[tbp]
\centering
\includegraphics{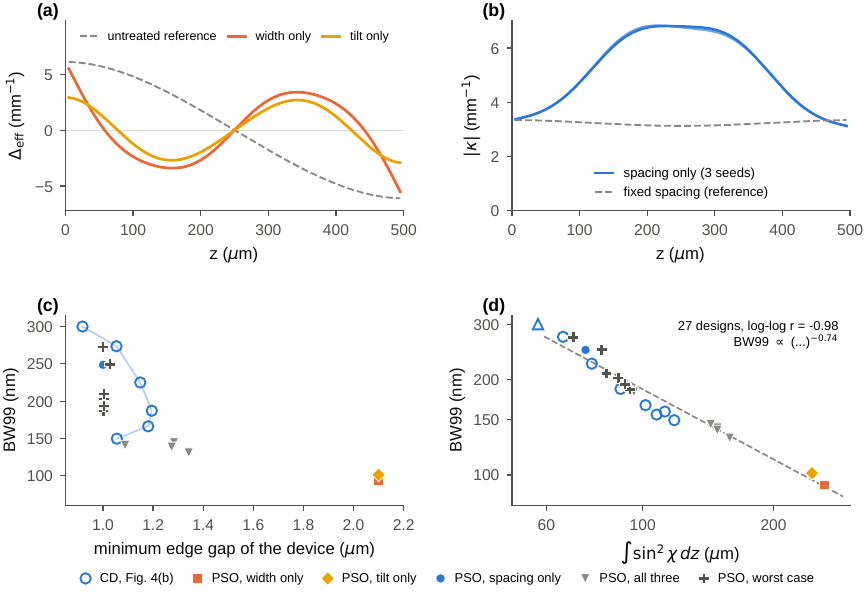}
\caption{PSO designs at $L=500$\um{} (Table~\ref{tab:3}). (a) Effective detuning $\Delta_{\eff}(z)$ of the width-only and tilt-only designs and of the untreated reference. (b) Coupling strength $|\kappa|(z)$ of the spacing-only designs compared with the fixed-spacing value. (c) BW99 against the minimum edge gap of the device and (d) against $\metric$, for the PSO designs (filled symbols) and the CD designs of Table~\ref{tab:2} and Fig.~\ref{fig:4}(b) (open circles); triangle: sweep edge; dashed: power-law fit to all designs below the sweep edge.}
\label{fig:6}
\end{figure}

We first run PSO for the width difference, the tilt, and the spacing alone, and for all three knobs, with $P_+$ at 1.55\um{} as the objective. For the width difference and the tilt, we use relaxed bounds (4\um{} for the width difference, 0.12~rad for the tilt, 300 iterations), because the default bounds limit the detuning that the width difference can supply. The results are summarized in Table~\ref{tab:3} and Fig.~\ref{fig:6}.

\begin{table}[tbp]
\caption{PSO with BPM in the loop at $L=500$\um. The first four rows maximize $P_+$ at 1.55\um{} with the knobs listed; the last two maximize, with all three knobs, the worst $P_+$ over the error range of Sec.~\ref{sec:platform} and over 1.45--1.65\um. Columns as in Table~\ref{tab:2}. Ranges are over seeds.}
\label{tab:3}
\centering
\small
\setlength{\tabcolsep}{4pt}
\begin{tabular}{>{\raggedright\arraybackslash}p{0.25\textwidth}ccccc}
\toprule
Knobs varied (objective) & \hd{Edge gap\\(\textmu m)} & $P_+$ & \hd{BW99\\(nm)} & \hd{Worst over\\errors} & \hd{$\metric$\\(\textmu m)}\\
\midrule
Width difference & 2.10 & 0.99985 & 93 & 0.9884 & 261\\
Axis tilt & 2.10 & 0.99948 & 101 & 0.9895 & 245\\
Spacing & 1.00 & 0.99998 & 249 & 0.9930 & 74\\
All three & 1.00--1.34 & 0.99979--0.99995 & 131--183 & 0.9914--0.9932 & 95--158\\
All three (worst case over error range) & 1.00 & 0.99968--0.99973 & 187--203 & 0.9935--0.9943 & 88--93\\
All three (worst case over 1.45--1.65\um) & 1.00--1.03 & 0.99911--0.99936 & 210--273 & 0.9928--0.9945 & 69--82\\
\bottomrule
\end{tabular}
\end{table}

All four reach $P_+\geq0.9995$, but they fall into two groups. The width-only and tilt-only designs converge to $\Delta_{\eff}(z)$ profiles of the same shape, with three resonance crossings [Fig.~\ref{fig:6}(a)], and to the same bandwidth of 93--101~nm. Their coupling strength stays at the fixed-spacing value (at most 3.43~mm$^{-1}$, against 3.34~mm$^{-1}$ for the reference). Width and curvature are two handles on one function, as Sec.~\ref{sec:knobs} predicts. The spacing is the only handle on the coupling strength, and the spacing-only designs raise it at the anticrossing [Fig.~\ref{fig:6}(b)], as the design rule does, and are 2.5 times broader. When all knobs are free, the optimizer mixes in detuning shaping, which is narrowband, and the bandwidth falls to 131--183~nm: more knobs do not give a broader device when the objective is set at a single wavelength.

Instead of $P_+$ at 1.55\um, we also maximize, with all three knobs, two robust objectives: the worst $P_+$ over the error range of Sec.~\ref{sec:platform}, which covers wavelengths from 1.50 to 1.60\um{} and the three fabrication errors applied one at a time, and the worst $P_+$ over 1.45--1.65\um. Robust objectives help. The first gives BW99 $=187$--203~nm, and the second gives 210--273~nm.

The searched designs follow the same metric as the constructed ones [Fig.~\ref{fig:6}(d)]. Shaping the detuning at fixed spacing leaves the field nearly perpendicular to the light near the middle of the device, as in a directional coupler ($\metric=245$--261\um), while boosting the coupling keeps the light close to the field (74\um). Over the 27 CD and PSO designs below the sweep edge, the correlation coefficient between $\log$ BW99 and $\log\metric$ is $-0.98$, and BW99 $\propto(\metric)^{-0.74}$. A uniform change of the precession rate in the two-mode model, averaged over the band, ranks these designs like BW99 (rank correlation $-0.97$), which supports the picture of Sec.~\ref{sec:bandwidth}.

None of the searched designs exceeds the design rule at the same minimum gap [Fig.~\ref{fig:6}(c)]. The spacing-only PSO gives 249~nm and the best robust design 273~nm, both at 1.00\um, the bound of the search, against 274~nm at 1.05\um{} for the design-rule coupler, whose worst $P_+$ over the error range (0.9948) is also higher than that of the designs that maximize it (0.9935--0.9943). Figure~\ref{fig:5}(d) shows this best robust design next to the design-rule coupler [Fig.~\ref{fig:5}(b)]. It uses all three knobs and is asymmetric: its narrowest gap, at the bound of the search, lies at $z=348$\um{} instead of mid-device, its tilt reaches 0.036~rad, about twice that of the design-rule coupler, and its width difference overshoots to $-2.0$\um{} near the output. Its BW99 matches that of the design-rule coupler, but its transfer at 1.55\um{} is lower, $P_+=0.9991$ against 0.9996 [Fig.~\ref{fig:5}(c)]. Each search costs 7200 BPM simulations, against a single construction.

\section{Discussion and conclusion}
\label{sec:conclusion}

Numerical optimization with BPM in the loop finds couplers with near-unity transfer for each knob alone and for all three, and in this sense it is a search over the same design space. The two-mode model tells what such a search has found: which function each knob serves, how far it steers the light from the field, and therefore how broad the result will be. The physics-based route reaches the same or a better result without a search, because it puts the coupling where it makes the reference adiabatic, leaves the CD correction small, and so keeps $\metric$ small. Numerical optimization remains useful for constraints outside the two-mode model, such as radiation loss in tight bends.

The present analysis has limits. The design study uses 2D scalar BPM on a low-contrast polymer platform without material dispersion. The 3D check of Appendix~C confirms its conclusions: rebuilt from the supermodes of the BCB channel waveguides, the three realizations again have the same bandwidth, the bandwidth follows $\metric$, and gap modulation again widens the band at a given minimum gap. The minimum gap, the length, and the bandwidth, however, change with the cross-section and its waveguide dispersion, and the 2D designs drawn in the 3D structure reach only $P_+=0.86$--0.97, so an exact design must be built from the supermodes of the actual structure. For high-contrast platforms such as silicon-on-insulator, the structure of Eqs.~\eqref{eq:KH}--\eqref{eq:Hz} carries over when the scalar overlaps are replaced by their full-vectorial counterparts, as discussed in~\cite{Lee2026}, and the design rule should be checked with full-vectorial simulations. The metric $\metric$ is motivated by the precession picture and verified on the designs of this paper; deriving its exponent from the dispersion of the coupling coefficients, and including that dispersion in the choice of the path~\cite{Stefanatos2026}, are natural next steps.

In conclusion, we have shown that mode-evolution couplers designed with width difference, spacing, and axis tilt belong to one design space. The three knobs act on the light through two functions, the coupling strength and the effective detuning: the tilt turns the phase of the coupling, and a rotation about $z$ moves that phase into the detuning, so the gauge, bent, and mixed realizations of CD driving are the same effective Hamiltonian and have the same bandwidth. BPM simulations show that the bandwidth follows the integral of $\sin^2\chi$ along the device, where $\chi$ is the angle between the light and the field on the Bloch sphere; for a CD design, this integral is set by the reference, not by the realization. The resulting design rule, to make the reference as adiabatic as the minimum gap allows, by narrowing its spacing at the anticrossing, and to let exact CD suppress the remaining nonadiabatic coupling, gives a 500\um{} coupler with ${>}99\%$ transfer over 274~nm without numerical optimization, which particle swarm optimization with any single knob or all three does not exceed. Three-dimensional BPM of the BCB channel waveguides confirms these conclusions, with the minimum gap set by the coefficients of the actual cross-section.

\section*{Appendix A. Accuracy of the two-mode model}
\addcontentsline{toc}{section}{Appendix A. Accuracy of the two-mode model}

\begin{figure}[tbp]
\centering
\includegraphics{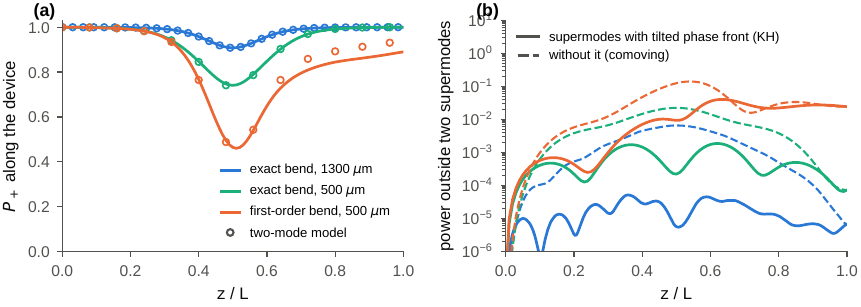}
\caption{(a) Supermode fidelity $P_+$ along $z$ from BPM (lines) and the two-mode model of Eq.~\eqref{eq:KH} (symbols) for the exact bent designs at $L=1300$ and 500\um{} and the first-order bent design at fixed spacing at $L=500$\um. (b) Power outside the two local supermodes along $z$, for supermodes with the tilted phase front (KH frame, solid) and without it (comoving frame, dashed).}
\label{fig:7}
\end{figure}

The two-mode model of Eq.~\eqref{eq:KH} reproduces BPM to within $4\times10^{-4}$ in $P_+$ at 1.55\um{} and $3\times10^{-3}$ at 1.50 and 1.60\um{} for all designs in this paper, and less than $10^{-4}$ of the power leaves the two supermodes at $L=1300$\um{} (Fig.~\ref{fig:7}). The exception is a large axis tilt at fixed spacing (0.056~rad at $L=500$\um), where 4\% of the power leaves the two-mode subspace. The model is accurate enough to rank designs and to construct them; all transfer and bandwidth values in this paper are from BPM.

Figure~\ref{fig:7}(b) also shows why the supermodes must carry the tilted phase front, as they do in the KH frame. Supermodes of the displaced cross-section without the tilt phase, as used in the lab and comoving frames, do not match the phase front of the light: at $L=500$\um, 2--5\% of the power appears outside them along the device, although it returns by the output. The two-mode truncation in the comoving frame therefore holds to first order in $kd\dot x_0$, the phase difference of the tilted front across the two waveguides, while the KH-frame truncation stays accurate for moderate tilts.

\section*{Appendix B. Construction of the exact CD couplers}
\addcontentsline{toc}{section}{Appendix B. Construction of the exact CD couplers}

\textbf{Tables.} We construct the CD couplers directly from the exact coefficients of Eq.~\eqref{eq:KH}, without fitting relations between geometry and coupled-mode parameters, in four steps. On a grid of width differences and spacings, we compute the supermodes at 1.55\um{} and tabulate the splitting $\Delta\beta$, the mixing-angle gradients $G_w=\langle\psi_-|\partial\psi_+/\partial(\delta w)\rangle$ and $G_d=\langle\psi_-|\partial\psi_+/\partial d\rangle$, and the position matrix elements $X_z$ and $X_x$ of Eq.~\eqref{eq:KH}. \textbf{Reference.} Along the reference path $\delta w(z)$, $d(z)$, the nonadiabatic coupling is
\begin{equation}
\frac{\dot\theta}{2} = G_w\,\dot{\delta w} + G_d\,\dot d,
\tag{B1}\label{eq:B1}
\end{equation}
and $\theta$ is obtained by integrating Eq.~\eqref{eq:B1} from the input, where $\theta$ is taken from the tables; the tables are anchored by mirror symmetry, $\theta=\pi/2$ at $\delta w=0$, and integrated along $\delta w$ at fixed spacing. The mixing angle depends on the path and not only on the end points, so it must be integrated along the actual path. The reference coefficients in the fixed basis are $\Delta_0=(\Delta\beta/2)\cos\theta$ and $\Omega_0=(\Delta\beta/2)\sin\theta$. \textbf{Target.} The CD Hamiltonian of Eq.~\eqref{eq:cd}, brought to a real coupling by the rotation about $z$ of Eq.~\eqref{eq:Hz}, gives the target coupling strength and effective detuning:
\begin{equation}
|\kappa| = \sqrt{\Omega_0^2 + (\dot\theta/2)^2}, \qquad \Delta_{\eff} = \Delta_0 - \frac{\dot\phi}{2}, \qquad \phi = \tan^{-1}\frac{\dot\theta}{2\Omega_0}.
\tag{B2}\label{eq:B2}
\end{equation}
\textbf{Geometry.} A fraction $f(z)$ of the coupling phase is supplied by the tilt,
\begin{equation}
\dot x_0 = \frac{f\,\phi}{k\,d_e},
\tag{B3}\label{eq:B3}
\end{equation}
where $d_e$ is the signed effective separation of the two waveguides: the difference of the centroids of the two fixed-basis states, $2\tilde X_z$ of Eq.~\eqref{eq:B4}, negative here because waveguide~1 lies at $x'<0$. Its magnitude reduces to $d$ for weakly coupled guides, so that $kd_e\dot x_0=f\phi$, the tilt phase of Eq.~\eqref{eq:tilt}. The width difference and spacing supply the rest. At each $z$ we solve for $\delta w$ and $d$ with Newton's method on the tables so that the effective Hamiltonian of the bent device, Eq.~\eqref{eq:KH} rotated back to the fixed basis, equals the target,
\begin{equation}
\frac{\Delta\beta}{2}\cos\vartheta - k\ddot x_0\tilde X_z = \Delta_{\eff}, \qquad \frac{\Delta\beta}{2}\sin\vartheta - k\ddot x_0\tilde X_x = |\kappa|,
\tag{B4}\label{eq:B4}
\end{equation}
where $\tilde X_z=X_z\cos\vartheta-X_x\sin\vartheta$ and $\tilde X_x=X_z\sin\vartheta+X_x\cos\vartheta$ are the position matrix elements in the fixed basis, and $\vartheta$ is the mixing angle of the device, integrated along its own path as in Eq.~\eqref{eq:B1}. Both sides have a real coupling: the KH transformation has moved the tilt into the curvature terms of Eq.~\eqref{eq:KH}, and the KH frame coincides with the lab frame at the ports, where $\dot x_0=0$, as the frame of Eq.~\eqref{eq:Hz} does where $\phi=0$. For weak coupling at fixed spacing, $|\tilde X_z|=d/2$ and $\tilde X_x=0$, and Eq.~\eqref{eq:B4} is Eq.~\eqref{eq:Hz} with the tilt phase $\phi=kd_e\dot x_0$. Where the spacing changes, Eq.~\eqref{eq:B4} has no term in $\dot d\dot x_0$: with $\dot x_0$ from Eq.~\eqref{eq:B3}, the curvature supplies $f\dot\phi/2-k\dot x_0\dot d_e/2$, and the width difference supplies the rest. In our bent and mixed designs, the second term reaches 14--26\% of the largest curvature term, and the curvature coupling $k\ddot x_0\tilde X_x$ stays below 1.5\% of $|\kappa|$. Because the realized device has its own path, and hence its own mixing angle, we repeat the solution with the path-integrated angle of the previous iterate until it converges, in 9--12 iterations to $3\times10^{-6}$~rad. The fixed bases of the device and of the reference coincide where their geometries do: at the ports for the bent realization, while for the gauge design they differ there by 0.11~rad, the port mismatch of Sec.~\ref{sec:three}. The construction takes about 0.1~s per design. For the gap-modulated references of Sec.~\ref{sec:rule}, the same steps apply along the path of Eq.~\eqref{eq:gap}.

\section*{Appendix C. Three-dimensional check}
\addcontentsline{toc}{section}{Appendix C. Three-dimensional check}

\begin{figure}[tbp]
\centering
\includegraphics{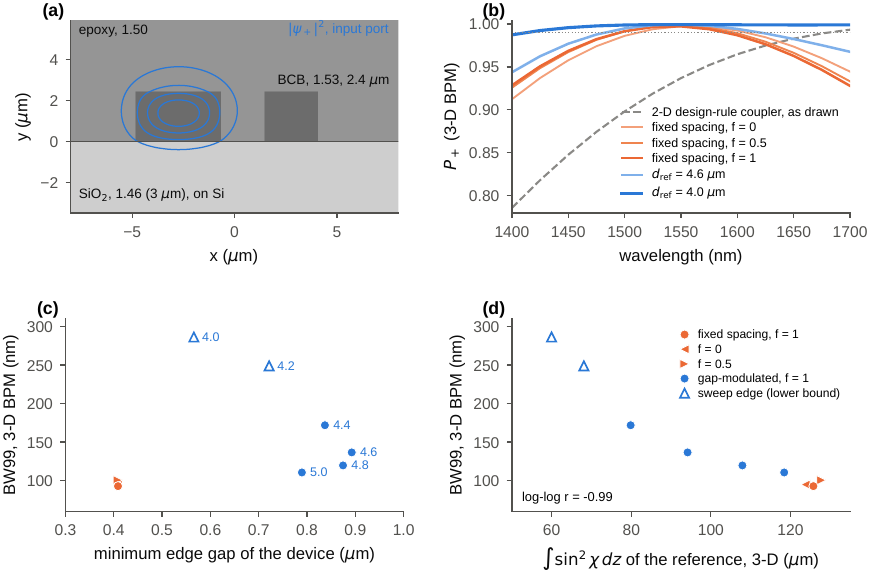}
\caption{Three-dimensional check on the BCB channel waveguides. (a) Cross-section at the input port ($W_1=4.2$\um, $W_2=2.6$\um, $d=5.5$\um) with contours of $|\psi_+|^2$. (b) $P_+$ against wavelength from 3D BPM at $L=500$\um: the 2D design-rule coupler drawn in the 3D structure (dashed) and exact CD designs built from the 3D supermodes, for the fixed-spacing reference in the gauge, mixed, and bent realizations and for gap-modulated references with $\dref=4.6$ and 4.0\um{} ($\sin^2$, $f=1$). (c) BW99 against the minimum edge gap of the device and (d) against $\metric$ of the reference, for exact CD designs at $L=500$\um, labeled by $\dref$ in \textmu m. Orange: fixed-spacing reference in the gauge, mixed, and bent realizations ($f=0$, 0.5, and 1; offset by $\pm2$\um{} in (d) for visibility); triangles: bands that reach the 1.70\um{} edge of the sweep (lower bounds).}
\label{fig:8}
\end{figure}

The 2D platform of Sec.~\ref{sec:platform} is the effective-index model of BCB channel waveguides~\cite{Tseng2013}: a 2.4\um{} thick BCB core ($n=1.53$) on a 3\um{} SiO$_2$ buffer ($n=1.46$) on a Si substrate, with an epoxy upper cladding ($n=1.50$) [Fig.~\ref{fig:8}(a)]. To check the conclusions in the actual structure, we simulate fully etched BCB channels using a scalar 3D finite-difference BPM solver with the alternating-direction implicit Crank--Nicolson scheme~\cite{Kawano2004} ($\Delta x=\Delta y=0.1$\um, $\Delta z=0.5$\um). The supermodes of each cross-section are computed by a two-dimensional finite-difference mode solver, and the exact coefficients of Appendix~B are tabulated from them. Refining the grid to $\Delta x=\Delta y=0.075$\um{} and $\Delta z=0.25$\um{} changes $P_+$ by at most $2.3\times10^{-4}$.

The 2D model reproduces the slab index but not the coupling. At $\delta w=0$, the 3D coupling is 0.74 times the 2D value at $d=5.5$\um{} and 0.98 times at 4.0\um, while the detuning at the input port is 36\% larger. The 3D coefficients also change faster with wavelength: from 1.45 to 1.65\um, the coupling at the anticrossing increases by 54\% in 3D and by 33\% in 2D.

The 2D designs, drawn in the 3D structure, therefore fail, with $P_+=0.86$--0.97 at 1.55\um{} [Fig.~\ref{fig:8}(b)]. An exact design is exact for the coefficients it is built on, so the coefficients must come from the actual cross-section. This applies equally to designs optimized with a 2D model in the loop.

We repeat the construction of Appendix~B with the 3D tables and verify each design by 3D BPM at $L=500$\um{} (Table~\ref{tab:4}). For the fixed-spacing reference, the gauge, mixed, and bent realizations reach $P_+=0.9971$--0.9984 and BW99 $=93$--100~nm, and all three narrow the edge gap to the same minimum of 0.41\um. The realization again does not set the band. The residual, mostly power outside the two supermodes (up to $2.6\times10^{-3}$), grows with the size of the CD correction. For the gap-modulated references, the correction is smaller and $P_+=0.9992$--0.9996.

\begin{table}[tbp]
\caption{Exact CD designs at $L=500$\um{} built from the 3D supermodes of the BCB channel waveguides and verified by 3D BPM ($\sin^2$ gap profile, bent realization $f=1$ unless noted). $\metric$ of the reference (Sec.~\ref{sec:bandwidth}). BW99 values marked $\geq$ reach the 1.70\um{} edge of the sweep. The first row is the 2D design-rule coupler of Sec.~\ref{sec:drc} drawn in the 3D structure.}
\label{tab:4}
\centering
\small
\begin{tabular}{lcccc}
\toprule
Design & \hd{Edge gap\\(\textmu m)} & $P_+$ & \hd{BW99\\(nm)} & \hd{$\metric$\\(\textmu m)}\\
\midrule
2D design-rule coupler, as drawn & 1.05 & 0.937 & -- & --\\
CD, fixed spacing, $f=0$ & 0.41 & 0.9973 & 95 & 126\\
CD, fixed spacing, $f=0.5$ & 0.41 & 0.9984 & 100 & 126\\
CD, fixed spacing, $f=1$ & 0.41 & 0.9971 & 93 & 126\\
CD, $\dref=5.0$\um & 0.79 & 0.9995 & 110 & 118\\
CD, $\dref=4.8$\um & 0.87 & 0.9996 & 120 & 108\\
CD, $\dref=4.6$\um & 0.89 & 0.9994 & 137 & 94\\
CD, $\dref=4.4$\um & 0.84 & 0.9993 & 172 & 80\\
CD, $\dref=4.2$\um & 0.72 & 0.9992 & $\geq249$ & 68\\
CD, $\dref=4.0$\um & 0.57 & 0.9992 & $\geq286$ & 60\\
\bottomrule
\end{tabular}
\end{table}

The bandwidth follows $\metric$ as in 2D [Fig.~\ref{fig:8}(d)], with a log--log correlation of $-0.99$ over the seven designs below the sweep edge. At the same $\metric$, the 3D bandwidths are lower than the 2D ones, for example 137~nm against about 190~nm at 94\um. We attribute the difference to the faster change of the 3D coefficients with wavelength, a larger $\delta$ in the picture of Sec.~\ref{sec:bandwidth}.

Gap modulation buys bandwidth at a given minimum gap in 3D as well [Fig.~\ref{fig:8}(c)]. The CD designs of the gap-modulated references have both a wider minimum edge gap, 0.57--0.89\um, and a larger bandwidth than the CD design of the fixed-spacing reference.

The absolute numbers differ from 2D. In 3D, BW99 $\geq249$~nm at $L=500$\um{} needs a minimum edge gap of 0.72\um, whereas in 2D the design-rule coupler gives 274~nm at 1.05\um{} (Table~\ref{tab:2}). The weaker coupling and the 36\% larger port detuning of the 3D structure make its reference less adiabatic at the same gap. The design rule carries over unchanged.

\section*{Funding}
{\raggedright National Science and Technology Council (\mbox{114-2221-E-006-049-MY3}, \mbox{115-2640-E-005-001}, \mbox{115-2640-E-005-002}).\par}

\section*{Disclosures}
The author declares no conflicts of interest.

\section*{Data availability}
Data underlying the results presented in this paper are not publicly available at this time but may be obtained from the author upon reasonable request.

\end{document}